\documentclass[aip,rsi,reprint,graphicx]{revtex4-1} 
\usepackage{amssymb}
\usepackage{graphicx}
\usepackage{amsmath}
\usepackage{gensymb}
\usepackage{braket}
\usepackage[version=3]{mhchem}

\begin{document}
\title{Fourier Domain Rotational Anisotropy Second Harmonic Generation}
\author{Baozhu Lu}
\affiliation{Department of Physics, Temple University, Philadelphia, PA 19122, USA}
\affiliation{Temple Materials Institute, Temple University, Philadelphia, PA 19122, USA}
\author{Darius H. Torchinsky}
\email{dtorchin@temple.edu}
\affiliation{Department of Physics, Temple University, Philadelphia, PA 19122, USA}
\affiliation{Temple Materials Institute, Temple University, Philadelphia, PA 19122, USA}




\date{\today}

\begin{abstract}
We describe a novel scheme of detecting rotational anisotropy second harmonic generation (RA-SHG) signals using a lock-in amplifier referenced to a fast scanning RA-SHG apparatus. The method directly measures the $n^{th}$ harmonics of the scanning frequency corresponding to SHG signal components of $C_n$ symmetry that appear in a Fourier series expansion of a general RA-SHG signal. GaAs was used as a test sample allowing comparison of point-by-point averaging with the lock-in based method. When divided by the $C_\infty$ signal component, the lock-in detected data allowed for both self-referenced determination of ratios of $C_n$ components of up to 1 part in $10^4$ and significantly more sensitive measurement of the relative amount of different $C_n$ components when compared with conventional methods.
\end{abstract}


\maketitle
\section{Introduction}

Rotational-Anisotropy Second Harmonic Generation (RA-SHG) has emerged as an effective tool for the study of crystallographic point group symmetry and the electronic symmetry breaking states of the surface and bulk~\cite{Fiebig2005,Kirilyuk2005,Mcgilp2010,Denev2011}. In the electric dipole approximation, second harmonic generation only arises from inversion symmetry breaking, thus applications of RA-SHG have commonly focused on ferroelectric order and acentric lattice and magnetic symmetry breaking. Recent experimental improvements~\cite{Torchinsky2014,Harter2015} have led to more ready application of the technique to single crystals and in cryogenic sample environments, enabling observation of lattice and electronic symmetry breaking that may be hidden to more conventional scattering probes~\cite{Torchinsky2015, Zhao2016,Harter2017,Zhao2017}, highlighting the technique's promise for investigating contemporary problems in condensed matter physics.

In a typical RA-SHG measurement, pulsed laser light is converted to its second harmonic frequency whose magnitude is measured as a function of rotation $\phi$ of either the polarization of the incoming and outgoing beams or between the scattering plane and crystalline axes. The resulting rotational anisotropy ``pattern" is then fit to a model where the nonlinear optical susceptibility encapsulates the point group symmetry of the crystal and any possible concurrent symmetry breaking electronic order. A significant impediment to observing subtle symmetry breaking using RA-SHG is the quality of the rotational anisotropy pattern obtained. To this end, a major advance inspired by low-noise approaches to THz polarization~\cite{Morris2012} and time-resolved reflectivity measurements~\cite{Elzinga1987, Edelstein1991, Feldstein1995} was the incorporation of fast-scanning into the RA-SHG technique~\cite{Harter2015}. While these improvements make the measurement insensitive to the low-frequency laser drift that plagues stop-start approaches, subtle alignment defects and imperfections in the apparatus optics can still lead to skewed and/or jagged RA-SHG curves. Extraneous signal sources can also arise from competing SHG channels as may occur, e.g., from heterodyning between surface second harmonic and a bulk quadrupolar response, resulting in asymmetric or otherwise flawed RA-SHG traces that obscure measurement of the desired signal. In other cases, the signal to noise ratio may simply be too small to precisely fit the data and reveal the presence of small relative changes in symmetry that may result from either a phase transition or laser excitation. These complications may be contrasted with more conventional diffraction-based scattering probes that exploit periodicity more directly, either reducing series of lattice planes into individual reciprocal space points of Laue patterns from single-crystals or generating sets of concentric rings in power diffraction patterns from polycrystalline or pulverized samples. This relationship between periodicity and singular measured values makes diffraction measurements relatively more insensitive to small defects or other imperfections in a manner that current RA-SHG data acquisition schemes do not.

In this paper we describe a novel method to detect RA-SHG signals and, in the spirit of diffraction probes, leverage their inherent periodicity as universally sinusoids as a function of rotational angle $\phi$. This method, built upon a fast-scanning RA-SHG spectrometer, is based on a Fourier decomposition of a signal by a lock-in amplifier's demodulators to select a single frequency component and phase of a periodic time-domain signal. When several concurrent frequencies are detected, one may be divided by the other to permit referencing of the various signal components to remove correlated noise. This scheme provides direct access to combinations of second order susceptibility tensor components and can result in up to a $10^2$ improvement in the signal-to-noise ratio of measurements of relative degrees of $C_n$ symmetry breaking, making the measurement more robust to signal imperfections, alignment defects and low signal-to-noise ratios as compared with currently used techniques.

\section{Background and Principle}~\label{Exp1}

In the electric dipole approximation, an electric field $E_i(\omega)$ of frequency $\omega$ incident upon a non-centrosymmetric crystal induces a radiated electric field proportional to the second order induced polarization $P_i(2\omega)$ through
\begin{equation}
P_i(2\omega)= \chi^{(2)}_{ijk}E_j(\omega)E_k(\omega)
\end{equation}
where $\chi^{(2)}_{ijk}$ is the second order optical susceptibility tensor reflecting the crystallographic point-group symmetry of the material through Neumann's principle~\cite{Birss1964}. Here we focus on RA-SHG measurements in which the crystalline axes rotate relative to the scattering plane although essentially identical arguments apply for all other RA-SHG geometries. In this geometry, the mathematical representation of rotating the crystalline axis by angle $\phi$ is given by transforming $\chi^{(2)}_{ijk}$ using the tensor
\begin{equation}
a_{ij} = \begin{bmatrix}
\cos(\phi) & -\sin(\phi) & 0\\
\sin(\phi) & \cos(\phi) & 0\\
0 & 0 & 1
\end{bmatrix}
\end{equation}
according to the standard transformation law for polar tensors
\begin{equation}~\label{eq:tlaw}
\chi'^{(2)}_{lmn} = a_{il}a_{jm}a_{kn}\chi^{(2)}_{ijk}. 
\end{equation}

In an experimental realization of this scattering geometry, the beam is introduced oblique to the sample surface allowing four different polarization combinations where the incident and/or emitted fields are polarized in $(P)$ or perpendicular $(S)$ relative to the scattering plane. The corresponding emitted intensities are denoted $I^{SS}_{2\omega}(\phi)$, $I^{SP}_{2\omega}(\phi)$, $I^{PS}_{2\omega}(\phi)$ and $I^{PP}_{2\omega}(\phi)$, where the first (second) superscript denotes the incident (emitted) polarization. In the absence of interfering signals sources that may mix - and hence heterodyne - with the $\chi^{(2)}_{ijk}$ response being studied, any given individual transformed tensor element $\chi'^{(2)}_{ijk}$ is thus at most proportional to the third power of a trigonometric function due to the transformation law Eq.~\ref{eq:tlaw} that requires one factor of $a_{ij}$ per tensor rank. Since field intensities are measured with a square-law detector, an arbitrary emitted intensity is given by $I^{MN}_{2\omega}(\phi)\propto|P_i(2\omega)|^2$ ($M,N=P$ or $S$), implying that the angular dependence of any RA-SHG data may be most generally expressed as
\begin{equation}\label{eq:shg-gen}
I^{MN}_{2\omega}(\phi) =  \left|\sum_{r=0}^3 \sum_{s=0}^{r} b_{rs} \cos^{r-s}(\phi)\sin^{s}({\phi})\right|^2,
\end{equation}
where the $b_{rs}$ represent linear combinations of individual tensor elements of $\chi^{(2)}_{ijk}$. The use of trigonometric product and reduction formulae allows Eq.~\ref{eq:shg-gen} to be recast as
\begin{multline}\label{eq:multordxy}
I^{MN}_{2\omega}(\phi)= X_0 + X_2\cos\left(2\phi\right) + X_4\cos\left(4\phi\right) + X_6\cos\left(6\phi\right)\\
+Y_2\sin\left(2\phi\right) + Y_4\sin\left(4\phi\right) + Y_6\sin\left(6\phi\right) 
\end{multline}
in a ``Cartesian" representation or, equivalently\begin{multline}
\label{eq:multord}
I^{MN}_{2\omega}(\phi)= A_0 + A_2\cos\left(2\phi +\psi_2\right)  
\\ + A_4\cos\left(4\phi+\psi_4 \right) + A_6\cos\left(6\phi+\psi_6 \right),
\end{multline}
in a ``polar" representation. In the above, the coefficients $X_n,Y_n,$ and $A_n$ comprise different combinations of the tensor components $\chi^{(2)}_{ijk}$ contributing signals of $C_n$ symmetry. The phases $\psi_n$ in Eq.~\ref{eq:multord} allow for rotational anisotropy components to be oriented arbitrarily relative to $\phi =0$ and may also be expressed in terms of the $\chi^{(2)}_{ijk}$. At all orders, these phases may have an offset that accounts for extrinsic tilt of the crystalline axes relative to the scattering plane if the sample axes themselves are tilted relative to the scattering plane at $\phi=0$.

Equations \ref{eq:multordxy} and \ref{eq:multord} constitute equivalent, exact Fourier expansions of a general dipolar homodyne RA-SHG signal. Odd orders of $\phi$ have been omitted as they may only appear in either a deliberately heterodyned experimental geometry or when more than one signal component is present, e.g., if a surface electric dipole signal interferes with the main bulk electric dipole contribution. In the absence of heterodyned signals, it is easily seen that each additional tensor rank of the susceptibility increases the order of the homodyne response by two; the nonlinear susceptibilities of second harmonic electric quadrupole radiation and third harmonic generation are both rank four, i.e., $\chi^{(2)}_{ijkl}$ and $\chi^{(3)}_{ijkl}$, respectively, thus the corresponding rotational anisotropy patterns include an additional term $X_8\cos(8\phi) + Y_8\sin(8\phi)$ in Eq.~\ref{eq:multordxy} or $A_8\cos(8\phi+\psi_8)$ in Eq.~\ref{eq:multord}.

In a fast scanned RA-SHG measurement, $\phi = 2\pi f_r t$ where $f_r$ is the rotational frequency of the experiment and $t$ is the elapsed time. The various signal components $X_n,Y_n,$ and $A_n$ manifest as sinusoids at frequencies $nf_r$ for $n\geq 1$ and are well suited to lock-in detection referenced to $f_r$ and its harmonics. Significantly, $n$-fold symmetries of the RA-SHG traces, representing intrinsic $n$-fold symmetry of the sample, appear on the $n^{th}$ demodulated harmonic response through the coefficients $X_n,Y_n,$ and $A_n$. As each term in Eq.~\ref{eq:multord} is determined directly and independently, the coefficients may be inverted to yield the values of individual $\chi_{ijk}^{(2)}$ tensor elements provided that the system is not underdetermined.

Recording the $X_n,Y_n,$ and $A_n$ values accurately using a lock-in amplifier requires the voltage input signal to be appropriately conditioned since the series of voltage spikes at the laser repetition rate produced from a photomultiplier tube or other optical detector contains Fourier components across the entire frequency spectrum in a complex manner depending upon the detector instrument response rather than only at signal frequencies $nf_r$. To address this issue, we may sample the peak voltage using a sample-and-hold (SAH) circuit that, upon triggering by a TTL pulse synchronized with the laser output, sustains the voltage amplitude until the next TTL pulse triggers the SAH to acquire the next peak value. The signal $s(t)$ is thus directly sampled at regular intervals to produce a continuous output where every laser pulse is sustained over a laser repetition period. Lock-in demodulation at frequency $f_r$ and harmonic $n$ is proportional to $\int \cos(2\pi n f_r t)s(t)dt$, effectively projecting out the $nf_r$ Fourier component of the piecewise constant function $s(t)$. When performed on the SAH output, this integration is akin to a Riemann sum as evaluated by the midpoint rule, with the error scaling inversely with the number of points, here the number of samples per revolution of the optics. 

The coefficients $X_0 = A_0$, however, cannot be demodulated referenced to $f_r$ as they represent the average value of a RA-SHG response $1/2\pi\int_0^{2\pi}  I^{FG}_{2\omega}(\phi)d\phi$. These terms can be recovered by mapping the signal onto a carrier frequency at the repetition rate of the laser $f_l$ and then demodulating the signal referenced to the laser output using a lock-in time constant $\tau\gg 1/f_l$, i.e., long relative to the rotational period of the optics. In practice, this is well executed using a carrier square wave so that correct values of $A_0$ may be obtained. The overall signal $s(t)$ is thus given by
\begin{multline}\label{eq:fd_sig}
s(t) = I_{2\omega}^{MN}(2\pi f_r t) \\
\times \left\{ \frac{1}{2}+ \sum_{k=1}^\infty\frac{2}{k \pi}\sin\left(\frac{k \pi}{2} \right)\cos\left(2\pi k f_l t\right)\right\},
\end{multline}
where $I_{2\omega}^{MN}(2\pi f_r t)$ is as represented in Eq.~\ref{eq:multordxy} or \ref{eq:multord} and the term in brackets is the Fourier expansion of an even square wave oscillating between 0 and 1. Examining the $k=1$ value of the expansion in Eq.~\ref{eq:fd_sig}, we see that $A_0$ may be recovered by demodulation at $f_l$ and multiplication by $\pi/2$ to account for the fact that the signal has been mapped onto a square wave. An additional factor of $\sqrt{2}$ is required in the event the lock-in provides a rms value rather than a peak-to-peak value. 

Signficiantly, we note that the electronic mechanism by which the pulses are sampled commonly leads to ``glitching," i.e., an instantaneous jump or spike of the voltage beyond the sampled value at each voltage change. This voltage overshoot corresponds to the Gibbs phenomenon known from Fourier analysis, which rescales every partial sum of the square wave expansion by a well known factor~\cite{Apostol1974}. Since we did not deglitch our SAH (which amounts to low-pass filtering of the signal), we multiplied the demodulated signal by an additional factor of 0.91 to account for the Gibbs overshoot.

In the above description, the signal modulates a carrier square wave at 5~kHz and the signal input to the lock-in effectively has a 50\% duty cycle.  As discussed above, the mathematical description of lock-in demodulation of these data is reminiscent of Riemann sums. It can be shown that setting every other term of a Riemann sum to zero yields approximately 1/2 the original value, and exactly 1/2 the original value of the integral in the limit~\cite{ATorchinsky}, which is suggested, albeit not proven, by the constant term in Eq.~\ref{eq:fd_sig}. The measured $f_r$ frequency input and all harmonics are thus, to an excellent approximation, demodulated as 1/2 of the value if there were no carrier wave meaning all measurements of $X_n, Y_n$ and $A_n$ $(n\geq 1)$ must be multiplied by 2 to account for the signal having support over 1/2 its domain. The signal must also be multiplied by $\sqrt{2}$ to convert from a rms to a peak-to-peak voltage.

Equation~\ref{eq:fd_sig} suggests additional demodulation schemes. The $n^{th}$ harmonic of the signal may be given as $\propto A_n\cos(2\pi n f_r t)\cos(2\pi k f_l t) = A_n/2 \left[\cos(2\pi (kf_l + n f_r) t) + \cos(2\pi (kf_l - n f_r) t)\right]$ showing that the $A_n$ are carried as the $n$ harmonic sidebands of the carrier frequencies $kf_l$. Provided that the appropriate reference frequencies can be generated, the $A_n$ thus may be directly demodulated at $kf_l \pm n f_r$ where each sideband accounts for 1/2 the spectral weight of the $A_n$ signal. Sideband demodulated signals must also account for glitching and rms voltage, as described above. Finally, we note that although it was not explored here, the signals may also be recovered by in tandem demodulation using two lock-ins.

The above description provides a faithful representation of the signal as emitted by the signal and is useful if the magnitudes of individual susceptibility tensor components are to be measured. We emphasize that the SAH may be omitted if such accuracy is not required as in, e.g., experiments designed to record changes in $C_n$ symmetry as a function of temperature rather than to deduce accurate values of the $\chi_{ijk}^{(2)}$.

\section{Experimental Setup}
As the RA-SHG apparatus will be described in detail elsewhere~\cite{Tran2018}, we only give essential details here, although any appropriately fast RA-SHG spectrometer, such as in Ref.~\cite{Harter2015}, may be used provided it can be interfaced with a lock-in. Briefly, the output of a 5~kHz repetition rate laser pumped an optical parametric amplifier (OPA) allowing for tunable incident energies over the range 0.48~eV - 2.58~eV. A 1.5~eV OPA beam from the second harmonic of the signal field was used in an all-reflective fast-scanning RA-SHG spectrometer whose kinetic optics were programmed to rotate synchronously at a $f_r=10$~Hz repetition frequency. The emitted SHG pulse was measured by a photomultiplier tube (Hamamatsu - R12829) biased by a high voltage power supply socket assembly (Hamamatsu - C12597-01), and the resultant current pulses converted to voltage pulses by a combination of charge integrator and shaper instrument (Cremat - CR-Z-PMT and CR-S-8us-US). For the gain settings used here, the detection electronics were calculated to produce an output of $\sim 260~\mu$V/photon at the detected 3.0~eV energy. In our RA-SHG spectrometer, each voltage pulse from the PMT is individually sampled by a fast data acquisition card (DAC - NI USB-6218) at its peak and recorded as a function of angle. Here the voltage pulses were first input into a sample-and hold circuit (SAH - Maxim DS1843) externally triggered at $2\times$ the laser repetition rate (10~kHz) generated by a combination of XOR circuit and digital delay generator (Highland Technology - T560) to alternately sample both the output voltage pulse and in between voltage pulses, each over a 300~ns window, thus modulating a 5~kHz square carrier wave by the signal.

The differential output of the SAH circuit was then connected to both the sampling input of the DAC and into the differential voltage input of a multi demodulator lock-in amplifier with AM/FM demodulation capability (Zurich Instruments MFLI + MF-MD + MF-MOD). The first lock-in reference was taken from a laser synchronous trigger at 5~kHz and used to demodulate the SAH output with a time constant $20~\text{s}\gg 100~\text{ms}$ (i.e, $\tau\gg 1/f$) at a filter slope of 48~dB/oct, providing a voltage proportional to the average value of the RA-SHG pattern to give $A_0$. The analog output generated by the DAC to coordinate the kinetic optics at 10~Hz repetition frequency also provided a sinusoidal reference signal to the second lock-in demodulator. This allowed synchronous measurement of the 5~kHz signal with one of the three possible harmonics of the 10~Hz RA-SHG electric dipole signal (i.e, at 20~Hz, 40~Hz, or 60~Hz) present in Eq.~\ref{eq:multord} (in the event that $A_0$ is not measured, all three harmonics can be recorded simultaneously). The same value of time-constant $\tau=20$~s was chosen for the 10~Hz demodulator as for the 5~kHz one. We also performed the experiment using simultaneous carrier wave and sideband demodulation at frequencies $f_l = 5$~kHz and $f_l-4f_r = 4960$~Hz, respectively, using the same value of $\tau$. 

Both the in-phase $X_n$ and out-of-phase $Y_n$ components were measured for all demodulated signals. The amplitude $A_n$ of the $n^{th}$ harmonic signal $A_n = \sqrt{X_n^2+Y_n^2}$ and the phase $\psi_n = \tan^{-1}\left(Y_n/X_n\right)$ were also directly recorded. In order to experimentally determine the uncertainty in the signal, we used a procedure similar to that implemented in hardware on the Stanford Research Systems SR830 lock-in amplifier and described in its manual~\cite{SRS}. After waiting for $\sim 20\tau$ for the voltage values to settle, a constant data stream was acquired for an additional $20\tau$ at a sampling rate of 837~samples/sec. The mean and standard deviation of the output signal were calculated to provide the measured values and their errors, respectively.

\section{Results and Discussion}

\ce{GaAs} was chosen as a prototypical test system due to widespread availability of high quality single crystal samples. A cubic crystal with acentric space group F$\overline{4}$3m (point group $\mathrm{T_d}$), the RA-SHG bulk electric dipole signals of GaAs can be easily computed to be 
\begin{align}
I^{SS}_{2\omega}(\phi)&= 0\\
I^{PS}_{2\omega}(\phi) &= 2\chi^2_{xyz}\left(1+\cos\left(4\phi\right)\right)\\
I^{SP}_{2\omega}(\phi) &= \chi^2_{xyz}/2\left(1-\cos\left(4\phi\right)\right)\\
I^{PP}_{2\omega}(\phi) &= \chi^2_{xyz}/2\left(1-\cos\left(4\phi\right)\right)
\end{align}
which, in princple, indicate that only the coefficent $A_0 = A_4$ is necessary to determine the full signal for any given polarization combination. In practice, the emitted RA-SHG signals from GaAs are significantly more complex due to interference from surface charging contributions~\cite{Germer1997}. We selected $I^{PS}_{2\omega}(\phi) $ since this response is four times stronger than the other responses and is significantly less influenced by the surface than the others at an incident photon energy of 1.5~eV. However, due to these extra signal contributions, we chose to model the data by two independent values $A_0+A_4\cos\left(4\phi\right)$.

\begin{figure}
    \includegraphics[width=0.45\textwidth]{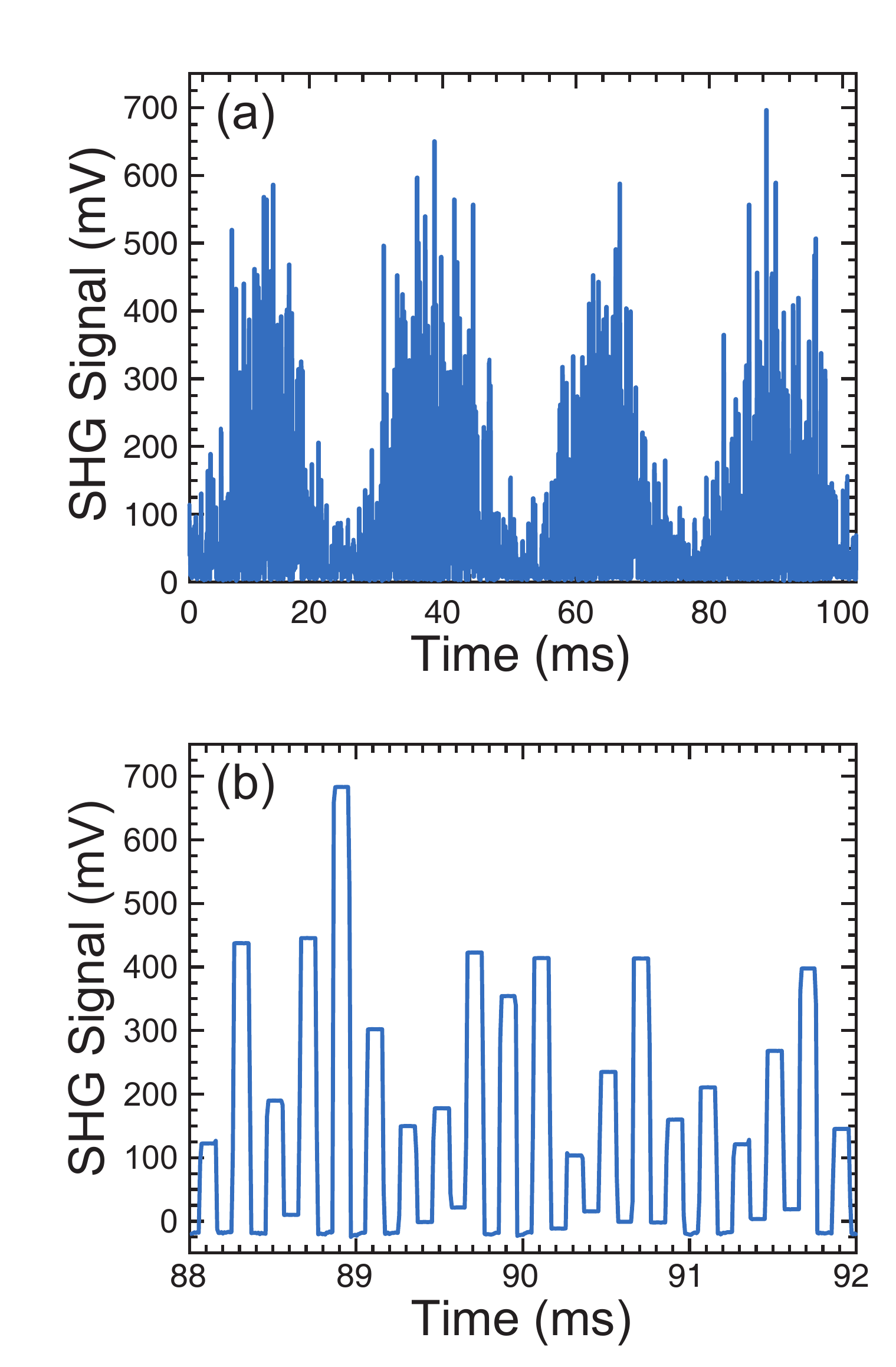}
    \caption{Raw data from the RA-SHG spectrometer. (a) On the timescale of one revolution of the kinetic optics, the four-fold symmetry of the GaAs response is observable through the four equally tall peaks over the 100~ms period of the experiment at frequency $4f_r$. This frequency may be demodulated to yield $A_4$. (b) At short times, the 50\% duty cycle output of the sample-and-hold hardware is evident, accounting for the ``filled-in" appearance of the data in panel (a). This $f_l=5$~kHz carrier frequency can be demodulated to yield $A_0.$ The sampling rate of the data are not sufficiently fast enough to see the effect of glitching.}
    \label{fig:raw_data}
\end{figure}

Plots of raw strong, high SNR signal data as input into the lockin from the SAH are shown in Fig.~\ref{fig:raw_data}a. The four-fold symmetry of the \ce{GaAs} $I^{PS}_{2\omega}(\phi)$ polarization geometry signal is evident in the panel from the four peaks of roughly equal height measured over the 100~ms period of revolution of the kinetic optics. The voltage oscillates between the emitted SHG intensity and a negligible background level due to the 5~kHz carrier square wave, which is more clearly seen in a zoomed-in portion of the data in Fig.~\ref{fig:raw_data}b.

\begin{figure*}
  \includegraphics[width=\textwidth]{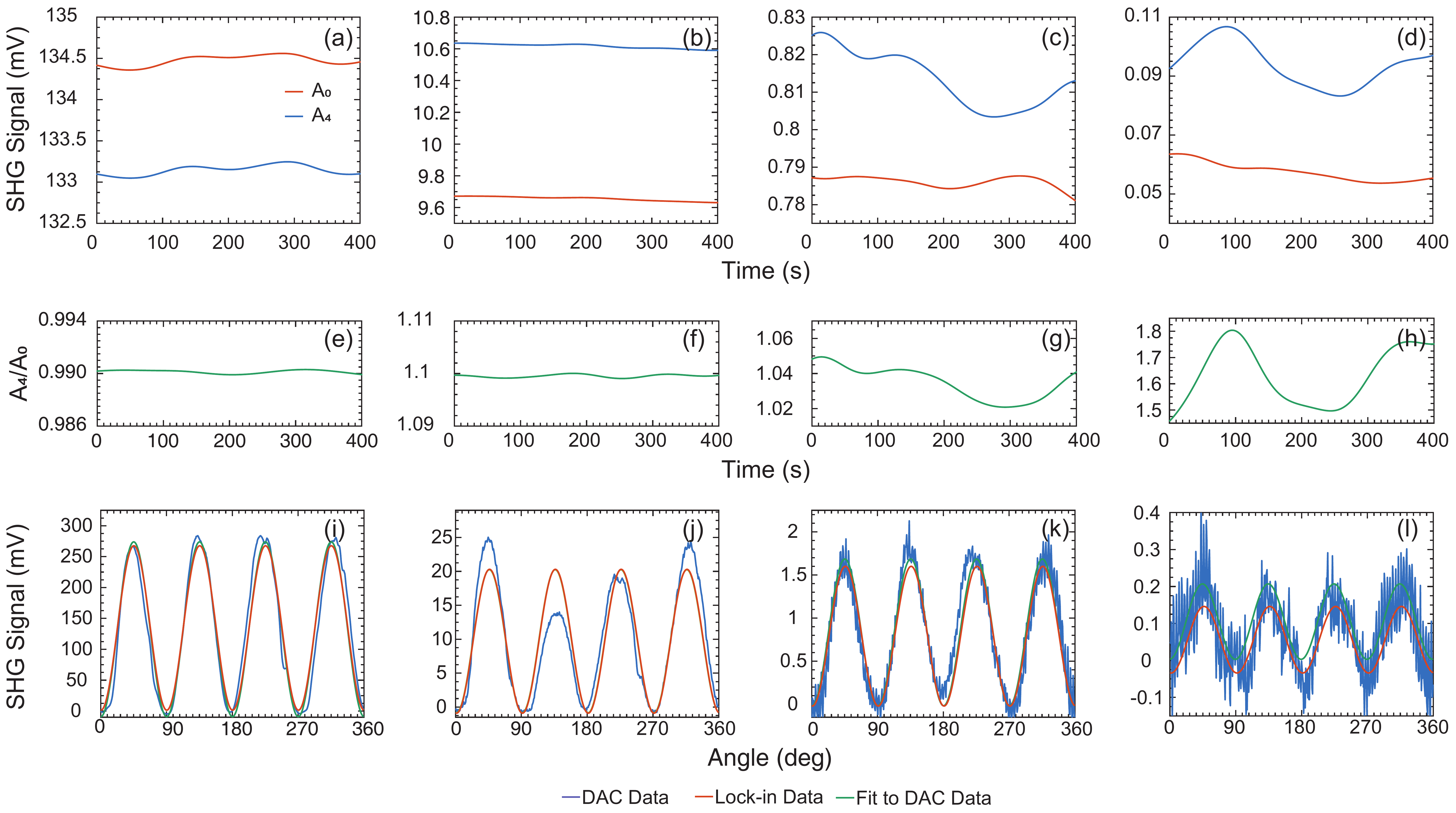}
  \caption{RA-SHG data as acquired by the lock-in amplifier and from the DAC. (a)-(c) show the the lock-in amplifier amplitudes of $A_0$ and $A_4$ as a function of time for $20\tau$ in various circumstances including: (a) well-aligned with good SNR, (b) poorly aligned with moderate SNR, (c) well-aligned with low SNR and (d) at the limit of experimental detection (on average $\lesssim 1$~photon/shot at maximum). The variation of the corresponding ratio $A_4/A_0$ is shown in panels (e)-(h). The DAC acquired traces are shown below the ratios in panels (i)-(l). Depicted in these panels are overlap of the DAC acquired raw RA-SHG data (blue), a reconstruction of the lockin data (red) and a line of best fit to the DAC data using the minimal model (green). The vertical scale is given in terms of a voltage output by the detection electronics with $\sim 260~\mu$V corresponding to the output voltage pulse equivalent to a single 3.0~eV photon.}
  \label{fig:demod}
\end{figure*}

RA-SHG data were acquired in four representative experimental configurations: well aligned with strong signal to noise ratio (SNR) ($\sim 1000$ detected photons/shot at measurement maximum on average); poorly aligned, i.e., a highly anisotropic plot as may arise from parasitic heterodyned SHG responses or samples with dirty surfaces at moderate SNR ($\sim 100$ detected photons/shot at maximum on average); well aligned with poor SNR ($\sim 10$ detected photons/shot at maximum on average); and well aligned at the limit of experimental detection ($\sim 1$ detected photon/shot at maximum on average). 

Data spanning these three decades of sensitivity are shown in Fig.~\ref{fig:demod} and the corresponding fit values are provided in Table~\ref{tab:example}. Panels (a)-(d) of the Figure show the output of the two demodulators as scaled by the appropriate numerical factors described in Sec.~\ref{Exp1} after subtracting an experimentally measured background with the laser blocked. The values of $A_0$ and $A_4$ are constant for large signal levels and appear to have larger degrees of variability as the strength of the overall signal is reduced, where this variability is more prominent in the $A_4$ component than the $A_0$ one at lower signal levels. From these time-series data, we determine the measured value of $A_n$ as the mean of its respective time-series data trace and the uncertainty in this mean as its standard deviation $\sigma$. We also computed the phase $\psi_4$ of the $A_4$ component from the lock-in time-traces of $X_4$ and $Y_4$ as $\psi_4 = \tan^{-1}\left(Y_4/X_4\right)$ in a similar manner.

\begin{table*}\caption{\label{tab:example}Comparison between fitted values and lock-in amplifier (LIA) measured values for the various configurations discussed in the text. $A_n$ are the magnitudes of the demodulated signals, $\psi_4$ is the angle of the $C_4$ component signal, and $r(A_0,A_4)$ is the Pearson correlation coefficient between the $A_0$ and $A_4$ components. Uncertainties are reported as the standard deviation of the LIA data or 67\% confidence interval for all types of fitted values.}
\begin{ruledtabular}
\begin{tabular}{lllll}
Coefficient  & LIA  & Fitted & Windowed & Sampled \\
\hline
\textit{High SNR}&&&& \\
\hline
$A_0$  & $134.50 \pm 0.06$~mV & $132.4 \pm 1$~mV & $132 \pm 3$~mV& $132 \pm 3$~mV\\
$A_4$  & $133.10 \pm 0.06$ mV & $142 \pm 2$~mV &$141 \pm 4$~mV  &$142 \pm 4$~mV  \\
$\psi_4$    & $-1.5976 \pm 3\times 10^{-4}$ &  $-1.58 \pm 0.01$ &$-1.73 \pm 0.03$&$-1.88 \pm 0.03$\\
$r(A_0,A_4)$ & 0.965 & - & - & -\\
$A_4/A_0$  & $0.9901 \pm 1\times 10^{-4}$& $1.07\pm 0.01$ &$1.07\pm 0.04$ &$1.07\pm 0.04$\\
\hline 
\multicolumn{5}{l}{\textit{Poor alignment/Moderate SNR}} \\
\hline
$A_0$   & $9.70\pm 0.01$~mV & $9.7\pm 0.1$~mV & $9.7\pm 0.4$~mV&$9.7\pm 0.4$~mV\\
$A_4$  & $10.60 \pm 0.01$~mV& $10.6\pm 0.2$~mV &$10.5\pm 0.5$~mV&$10.5\pm 0.5$~mV\\
$\psi_4$  & $-1.677 \pm 0.001$ & $-1.66\pm 0.02$ &$-1.81\pm 0.05$ &$-1.96\pm 0.05$  \\
$r(A_0,A_4)$ & 0.982& - & - & - \\
$A_4/A_0$  & $1.0995 \pm 3\times 10^{-4}$& $1.09 \pm 0.02$ &$1.09 \pm 0.07$  &$1.09 \pm 0.07$\\
\hline 
\textit{Low SNR} \\*
\hline\nopagebreak
$A_0$  & $790 \pm 1~\mu$V & $830 \pm 10~\mu$V & $840 \pm 20~\mu$V & $840 \pm 20~\mu$V\\*
$A_4$ & $810 \pm 7~\mu$V & $850 \pm 10~\mu$V & $840 \pm 30~\mu$V & $850 \pm 40~\mu$V\\*
$\psi_4$  & $-1.65 \pm 0.01$ &$-1.60 \pm 0.01$& $-1.77 \pm 0.03$ &$-1.89 \pm 0.04$ \\*
$r(A_0,A_4)$ & 0.200  & - & - & -\\*
$A_4/A_0$  & $1.04 \pm 0.01$ &$1.02 \pm 0.02$ & $1.01\pm 0.04$ & $1.02\pm 0.05$\\*
\hline
\multicolumn{5}{l}{\textit{Detection Limited}} \\
\hline
$A_0$  & $58 \pm 3~\mu$V & $76 \pm 4~\mu$V &$76 \pm 6~\mu$V&$80 \pm 9~\mu$V\\
$A_4$ & $97 \pm 7~\mu$V & $49 \pm 6~\mu$V & $135 \pm 9~\mu$V& $110 \pm 10~\mu$V\\
$\psi_4$  & $-1.67 \pm 0.08$ & $-2.4 \pm 0.1$ &$-1.38 \pm 0.06$& $-1.9 \pm 0.1$\\
$r(A_0,A_4)$ & 0.486  & - & -& -\\
$A_4/A_0$  & $1.7 \pm 0.1$ & $0.65\pm 0.08$ &$1.8\pm 0.2$ & $1.4\pm 0.2$ \\
\hline
\multicolumn{5}{l}{\textit{Sideband Demodulated, no SAH}} \\
\hline
$A_0$  & $34.7 \pm 0.7~\mu$V & $221\pm 9~\mu$V & $220\pm 10~\mu$V & $220 \pm 20~\mu$V\\
$A_4$  & $34.7\pm 0.6~\mu$V & $230\pm 10~\mu$V  & $220\pm 20~\mu$V &$230 \pm 30~\mu$V  \\
$\psi_4$    & $-1.62 \pm 0.04$ &$-1.61 \pm 0.05$  & $-1.71 \pm 0.08$ &$-2.1 \pm 0.1$\\
$r(A_0,A_4)$ & -0.37 & - & - & -\\
$A_4/A_0$  & $1.00 \pm 0.03$& $1.01\pm 0.07$ & $1.0\pm 0.1$ &$1.0\pm 0.2$ \\
\end{tabular}
\end{ruledtabular}
\end{table*}

Simultaneous to acquiring the lock-in data, we performed a standard point-by-point average of the SAH voltage output signal during the exact same time period using the DAC. These data are plotted in Figs.~\ref{fig:demod}(i)-(l) after subtracting a separately measured background signal when the laser was blocked. We note that for the DAC-sampled data at the detection-limited voltage level, this background had to be estimated due to the stated absolute accuracy of the DAC. Superposed with the data are reconstructions of the RA-SHG signals using the parameters from the ``LIA" column of Table~\ref{tab:example} as well as fits to the DAC data in the MATLAB environment using the function $s(\phi) = A + B\sin\left(4*\phi + \psi\right)$. The fit values with errors determined as 67\% confidence interval are given in Table.~\ref{tab:example}. There is excellent agreement between all three quantities, indicating that the factors discussed in Sec.~\ref{Exp1} faithfully replicate the data.

\begin{figure}
    \includegraphics[width=0.45\textwidth]{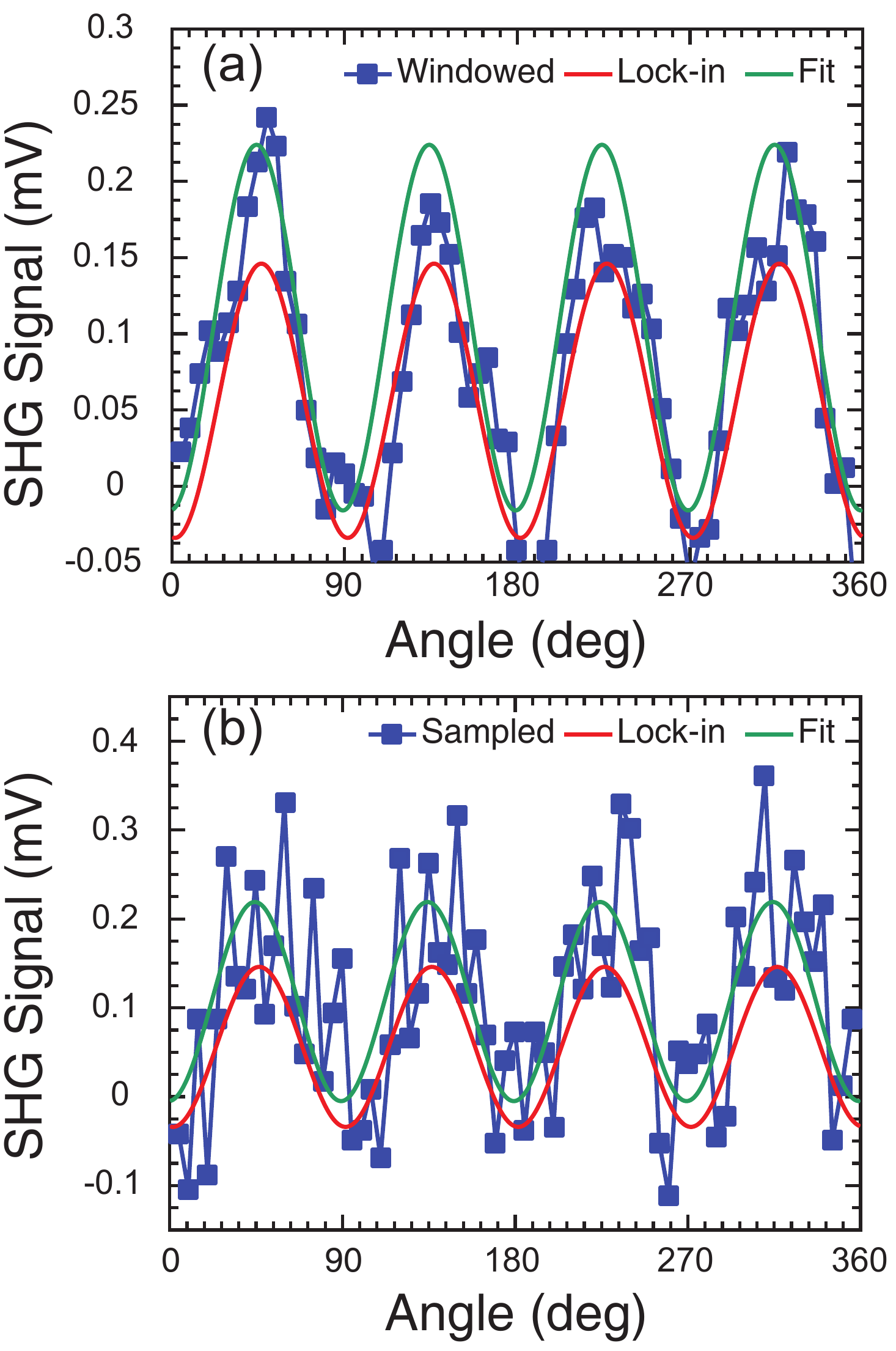}
    \caption{Comparison of lock-in with ``data-sparse" traces representative of most currently used RA-SHG detection schemes for signals at the threshold of experimental detection. The data from Fig.~\ref{fig:demod}(i) are either (a) averaged over a window every $5\degree$ or (b) sampled every $5\degree$. Also shown are fits to each data acquisition scheme and the reconstructed lock-in signal using the data of Table~\ref{tab:example}.}
    \label{fig:sparse}
\end{figure}

The ``data-dense" traces of Figs.~\ref{fig:demod}(i)-(l) comprise 503 data points per revolution and are atypical for most other RA-SHG measurements which are more ``data-sparse" by roughly an order of magnitude. In order to compare the method with more common, ``data-sparse" RA-SHG techniques, we include a plot of a ``windowed" trace in which our fast scanned data are averaged over a $5\degree$ interval, as well as a ``sampled" measurement in which a data point is selected from every $5\degree$. Plots of these data with their fits are shown in Fig.~\ref{fig:sparse} alongside the reconstructed lock-in signal. The corresponding fit values and errors are also compiled in Table~\ref{tab:example}.

Several trends emerge from examining the data for $A_0, A_4$ and $\psi_4$ in the Table~\ref{tab:example}. The lock-in based data generally have smaller error bars, with the relative error increasing as the signal strength decreases. Comparing ``data-dense" and ``data-sparse" fits, there is a further increase in the uncertainty of the fitted values with a reduction in data density. These findings not only emphasize the difference between commonly used RA-SHG data acquisition techniques and the lock-in based method, but also the poorer precision obtained with fewer data points. Significantly, we note that the uncertainty in the lock-in acquired values varies by roughly one order of magnitude, from at most $60~\mu$V for the strongest signals to least $3~\mu$V for the weakest, never representing an average change of more than a fraction of a single photon, i.e., $260~\mu$V , on average. The range in uncertainty of the DAC based data varies by three orders of magnitude over the three signal strength decades of the measurement, demonstrating the lock-in measurement to be considerably more precise, particularly at higher SNR levels. The only parameter range in which the lock-in and DAC-based methods show marked disagreement is at the limit of experimental detection comprising, on average, a single photon per shot at the maximum of the RA-SHG trace. We believe this is due to error from the SAH as described below.

Noise in the demodulated signal derives from a combination of slow laser drift, the noise spectrum of both the detection electronics and the laser, and the signal shot noise that emerge through the passband of the lock-in amplifier filter. Except for laser drift, all these noise sources are incoherent between different parts of the frequency spectrum and thus produce uncorrelated drifts in the $A_n$. By contrast, laser drift affects the amplitude of the signal as a whole and thus causes a uniform and simultaneous change in the magnitudes of all frequencies, i.e., in the $A_n$. In order to discriminate between correlated and uncorrelated sources of noise, we computed the Pearson correlation coefficient $r(A_0,A_4)$ of the two vectors of measured data points for $A_0$ and $A_4$ as plotted in Figs.~\ref{fig:demod}(a)-(d). Correlation values $r(A_0,A_4)\approx 1$ indicate that the noise derives exclusively from laser drift, while $r(A_0,A_4)\lessapprox 0$ specifies an incoherent, electronic origin. In the case of a large positive value of $r(A_0,A_4)$, one time-series data set may thus be divided by the other to yield an effectively self-normalized and self-referenced representation of the data. The ratio $A_4/A_0$ may then be computed by calculating the mean of this point-by-point division, while the error is computed as its standard deviation. Plots of the ratios for the four representative signal configurations are shown in Figs.~\ref{fig:demod}(e)-(h). The corresponding quantities for the various kinds of fitted values are obtained by dividing the fitted values for $A_4$ by that those for $A_0$ and using standard error propagation to obtain the error. 

The calculated correlation coefficients and $A_4/A_0$ values obtained by both division of the lock-in data and fitting to the various forms of DAC acquired data are all given in Table~\ref{tab:example}. We note that at the highest signal amplitudes probed, $r(A_0,A_4)\approx 1$, indicating that essentially the entirety of the error in the signal is due to laser fluctuations and the corresponding error in the ratio $A_4/A_0$ is 100 times lower than the next best data acquisition method, the ``data-dense" fitted values. When compared with the ``data-sparse" methods representative of all other RA-SHG apparatus, the error is 400 times less.

The correlation coefficient generally drops with the SNR, indicating that increasing amounts of noise come from electronic sources as signal strength diminishes. This is to be expected as the SHG experiment is a dark-field, homodyne measurement: excepting shot noise, the intrinsic optical noise will constitute the same percentage of the total signal at all emitted SHG levels, making electronic noise progressively more prevalent as light levels diminish. We attempted to improve the measured SNR of the $A_4/A_0$ ratio in the lock-in based measurement through the adjustment of several experimental parameters, but neither greater amplification by the PMT nor higher gain in the detection electronics changed the degree of relative noise and relative error in the ratio. Similarly, the use of longer time constants and/or accumulating data for a larger number of time constants did not make a discernible difference on the error in any of the measured quantities beyond extending the measurement duration to $40\tau$. We did not observe a demonstrable improvement in relative error for the ratio when using sideband demodulation for various signal levels, but noticed a decrease in absolute error in $A_4$ relative to $A_0$ along with a decrease in correlation coefficient $r(A_0,A_4)$.

We remark that values from the detection limited case are considerably less accurate than those obtained with stronger signals. We posit that this derives in part from error in the SAH electronics at extremely low voltage levels, but also from the stated accuracy limit of the DAC. Thus, we have also taken data near the detection limit threshold in a configuration without the SAH. In order to require the fewest amount of multiplicative correction factors, we took these data using the AM demodulation option of the MFLI lockin, referencing the two demodulators to the fundamental ($k=1$) carrier wave at $f_l = 5$~kHz and to the difference between this carrier and signal frequency at $f_l - 4f_r = 4960$~Hz. An experimentally measured background was also subtracted from the signal. As expected, we observed a large discrepancy between the absolute values as measured by the lock-in and the DAC as the Fourier coefficients of an expansion like that in Eq.~\ref{eq:fd_sig} now depend on the instrument response function of the detection electronics. However, we recover excellent agreement of the value $A_4/A_0$ as compared with the high SNR case, with the lock-in based method again representing the smallest relative uncertainty by roughly a factor of 2 to the next cleanest method. We note that for the lowest light levels measured, $r(A_0,A_4)$ was negative for sideband demodulation, indicating completely uncorrelated noise and requiring that the ratio be computed according to standard error propagation.

\section{Conclusions}

We have shown that a RA-SHG signal may be exactly represented as a finite Fourier series in factors of the rotational angle $\phi$ of $C_n$ symmetry and have described a method of interfacing a fast RA-SHG spectrometer with a lock-in amplifier to directly measure these Fourier coefficients. An example of our technique applied to single crystal \ce{GaAs} indicates that our lock-in based method performs better than competing RA-SHG measurement techniques for a wide range of SNR values, particularly at higher SNR, i.e., $\gtrsim 1000$~photons/shot on average. Our results also show that even at moderate to low SNR levels, the correlated nature of the noise allows for time-series measurements of different components of $C_n$ symmetry to be divided by one another to deduce their ratios to significantly more sensitive levels than was accomplished with signal averaging and fitting. As the various frequency components corresponding to different terms in the expansions of Eqs.~\ref{eq:multordxy} and \ref{eq:multord} can be measured simultaneously and independently, we propose that our method will allow for sensitive detection of changes in crystalline or electronic symmetry even in the absence of ``well-aligned" signals.

Having determined that our primary source of noise is likely electronic, we suspect that the use of cleaner electronics than those we have used here, in particular a more low-noise, deglitched SAH, would allow more precise and accurate measurement. The PMT used here could also be replaced by one that is cooled to reduce the dark current. A faster repetition rate laser could be used along with a more stable mechanism of increasing $f_r$ to decrease $1/f$ noise components and averaging time. However, since the lock-in method is clearly superior to any form of sampling measurement at moderate SNR levels and above ($\gtrsim 100$~photons/shot), we suggest that further experimental improvements focus on enhancing the detected photon yield of the SHG above this limit. Simple experimental corrections, such as using larger spot sizes or shorter pulse durations, can be used to preserve the same sample fluence (and avoid sample damage) while yielding more SHG photons per laser shot. For small samples and systems employing high numerical aperture objective optics, this solution may not be practical, meaning schemes employing heterodyne detection~\cite{yazdanfar2004, Wilcox2014} or stimulated SHG~\cite{Goodman2015} should be investigated. Such optical signal amplification techniques will allow for measurement of additional signal components that are buried in the noise and difficult to resolve, with the primary limitations being the number of demodulators that simultaneously analyze the signal. Finally, we note that this method could be interfaced to time-resolved pump-probe measurements~\cite{Shank1983} to detect similarly small photoinduced changes in symmetry, as well as time-domain measurement of impulsively driven collective modes~\cite{Chang1997} with a potential resolution approaching $\Delta A_n/A_n\approx  10^{-4}$.

\begin{acknowledgements}
DHT acknowledges Temple University start-up funds.
\end{acknowledgements}
\bibliography{SHGFD}

\end{document}